# Effects of Interfacial Distance and Electric Field on Graphene-Silicene Hybrid Structures

K M Abeywickrama, P K D D P Pitigala, W W P De Silva*

*University of Sri Jayewardenepura*

*wasanthidesilva@sjp.ac.lk

## ABSTRACT

Graphene is a two-dimensional (2D) semimetal with high mobility in charge carriers due to the existence of Dirac points. Silicene is another promising material, with properties analog to graphene. Many silicon (Si) based electronic devices can be integrated via graphene-silicene (Gra/si) hybrid structures. These electronic applications are mostly based on the ability of tuning the band gap via electronic structure deformation that is expected to be achieved by multilayer stacking, applying transverse external electric field (EF) and altering interfacial distance. In this work we investigate the band structure, density of states (DOS) distribution and the band gap with respect to interfacial distance and transverse external EF for the unit cells of bilayer graphene, monolayer silicene, andhybrid $Si_2C_6$. First principle calculations were carried using Quantum espresso software based on the Density Functional Theory (DFT).

## 1. INTRODUCTION

The two-dimensional (2D) materials have gained greater interest among the science community due to unusual properties identified for the graphene. Graphene is a 2D structure made out of carbon atoms arranged on a honeycomb structure. Graphene was a significant investigation with ground breaking experiments, which is awarded by the Nobel Prize in Physics in the year 2010 to the two Physicists, Sir Andre Geim and Sir Kostya Novoselv. They were extracting one-atom thick graphene from graphite using the sticky Scotch tape[1]. The tape is used to peel off thin layers from graphite until they obtained the one-atom thick material of carbon, the graphene.

Carbon, the fundamental atom of graphene, is the fascinating element in organic materials, which are showing important structural, electrical, optical and mechanical properties due to their flexible bonding nature in unlimited structural arrangements. Carbon atom has $1s^2 2s^2 2p^2$ electronic configuration with 4 electrons in the outer shell. In graphene, one carbon atom formed $\sigma$ bond with the three carbon atoms due to $sp^2$ ($sp_x$ and $sp_y$) hybridization and leave one electron in the $p$ orbital perpendicular to the $xy-$ plane. That $p_z$ orbital consist of highly-mobile electrons ($\pi$ electrons) responsible for the formation of covalent bond with another neighboring carbon atom. The bonding and anti-bonding (conduction and valence) $\pi$ bands form six zero gap points in the Brillouin zone which are called as the Dirac points[2].At the Dirac point, the effective mass of the electrons is zero, and the energy follows the linear dispersion relation ($E = hV_F k$) with respect to the momentum, which leads to high charge carrier mobility.

Due to these effects, graphene can be identified as a semi-metal material found with the best conductivity at room temperature due to massless electrons and Dirac node. Isolated graphene is less productive due to their uncontrollable charge flow, and it will limit the use of pure graphene in electronic applications. Hence, it requires methods to regulate the charge flow in graphene, by altering the energy band structure.

Silicene consists of Si atoms which is analog to the graphene one-atom think 2D hexagonal honeycomb lattice structure as shown in figure 5-(a). Silicene has low buckled structure due to $sp^2$ and $sp^3$ hybridization unlike planar graphene as shown in figure 5-(c). First, theoretically predicted the silicene by Takeda and Shiraishi in 1994[3] and later synthesis by Guzman-Verri and Voon 2007. It has $sp^2$ hybridize Si atoms, with linear electronic dispersion relation at the Dirac point like grapheme [4].

Gra/si hybrid structures gives promising materials for Si based electronic devices. There are many reasons for selecting silicene as the substrate for graphene. Few of them are hexagonal symmetry, similar lattice constants, Dirac cone shape of zero band gap, high carrier mobility, and tunable band gap with respective transverse EF. The hybrid structures can be formed using several combinations as an example $SiC_2$, where the Si atom is placed above the center of each hexagon in graphene, and $Si_nC_m$, where n and m are integers. There is some work that needs to be identified the properties for individual structure to investigate the most reliable hybrid system.

There are many possibilities to regulate the band structure/gap; applying a transverse external electric field[5], changing interfacial distances[6], doping and hydroganization [7] are few of them. The controllability of the band gap in graphene opens up the opportunity to use it in many applications such as construction of Nano-electronic devices[8], optoelectronic devices, radio frequency devices[9], sensors[10], transparent electrodes for solar cell [11], energy storage and other aspect of mechanical reinforcement.

For the first time, a bilayer graphene configuration has been suggested as a way of band gap opening by McCann and Falko [12]. Here, the band gap in the graphene bilayers structure opens up due to symmetry breaking of the honeycomb structure. The van der Waals interaction between the layers, will induced a potential difference, creating a gate between the layers and portion of electronic states will increased their energies by $+\Delta$ and the other electronic states willdecreased their energies by $-\Delta$, which leads to the existence of $2\Delta$ band gap in bilayer graphene.

Changing the interlayer distance can tune the Van der Wal interaction between layers. At smaller distances a high interaction between the layers leads to open up the band gap near Dirac point. For large interlayer distance, weaker the interaction and hence decrease the band gap. Two layers tend to show their intrinsic properties after exceeding their critical interfacial distance, hence a graphene bi-layeracts as monolayer of graphene for large distances. Similar, effects can be observed as well. The hybrid structures of Gra/si are more interesting, and they have a possibility to form the p-n junction, metal-semiconductor (Schottky junctions) and reversing p-n junction, as well as conductors due to changing the interfacial distance.

Bilayer graphene would create an interlayer potential asymmetric under the transverse electric field, which induces the band gap. Therefore, the band gap is increased due to an external electric fieldforgraphene, silicene and Gra/si hybrid structures.

## 2. COMPUTATIONAL METHOD

The first-principle calculations have been performed using open source Quantum expresso (QE) 6.4.1 .The calculations were carried out in plane wave basis set within the Local Density Approximation (LDA) which represents the exchange correlation effects. C.pz-vbc.UPF and Si.pz-vbc.UPF were used as pseudo potentials for C and Si, which count the effect of valence electron contribution to the calculation. First, geometrical optimization is done using the relaxation technique for the cell parameters and atomic positions of the system (Figure 1-(a)). The energy cut-off for the plane wave is obtained using total energy convergence as shown in figure 1-(b).The selecting K-path and sampling K-points are the most important in the band structure calculation. All the geometries are followed hexagonal structure throughout this work, hence, we are interested on hexagonal structure, which has highly symmetric k-path via $\Gamma \rightarrow K \rightarrow M \rightarrow \Gamma$ as shown in figure 2.The XCrySDen software was used to visualize the atomic positions and select the k-path in order to get the appropriate number of k-points. We select the appropriate k-mesh values as shown in figure 1-(c). The self-consistent calculations and total energy calculations were performed for graphene, silicene and $Si_2C_6$.

Band structure plots are the pictorial representation of the energies for k-points within the first brillouin zone inside the material. In these band structures Fermi energy ($E_f = 0$) is considered as a reference line and other band lines are taken using $E - E_f$. The X-axis represents the k-points along the highly symmetric k-path of the two dimensional hexagon brillouin zone for the graphene, silicene, and Gra/si combination ($Si_2C_6$). Density of state plots are visualized the number of electronic states for a given energy in k-path.

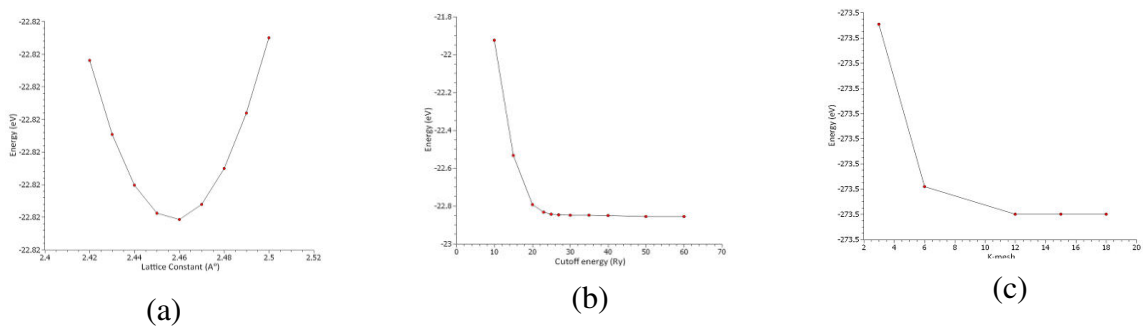

(a)    (b)    (c)

**Figure 1:** Optimized (a) lattice constant (b) cut-off energy (c) k-mesh for unit cell of graphene for QE calculation.

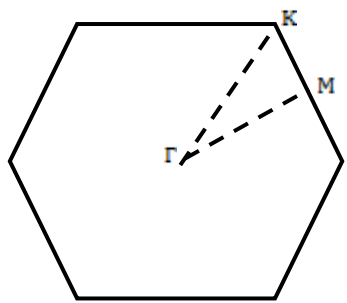

**Figure 2:** The first Brillouin zone for graphene. K-path define as $\Gamma \rightarrow K \rightarrow M \rightarrow \Gamma$.

## 3. RESULTS AND DISCUSSION

### 3.1 Graphene

Figure 3-(a) shows unit cell of graphene with the optimized lattice constants $a = b = 2.4352735\ A^o$. Other optimized parameters, vacuum = $20\ A^0$, cut off energy =30 Ry , k-mesh = 12 12 1 and k-point 70, are used for the DFT calculation. First, obtain the band structure and DOS plot for the unit cell of graphene as shown in figure 4-(a). The Dirac point exists at K and DOS plot is aligned with the band structure behavior. This implies how accurate our calculation for further studies. According to the figure 4-(b), bilayer graphene indicates the gap at K-point point due to mirror symmetry of these two layers. In the calculation of bilayer graphene, interlayer distance keep as 3.35 $A^0$ and cut off energy 25 Ry. Bilayer graphene is useful configuration for Field Effect Transistors (FET), which can be achieved by tuning band gap under the doping[13] and applying EF[14]. Figure 4-(d) indicates how tunable the band gap with respect to external EF perpendicular to the graphene bilayers.

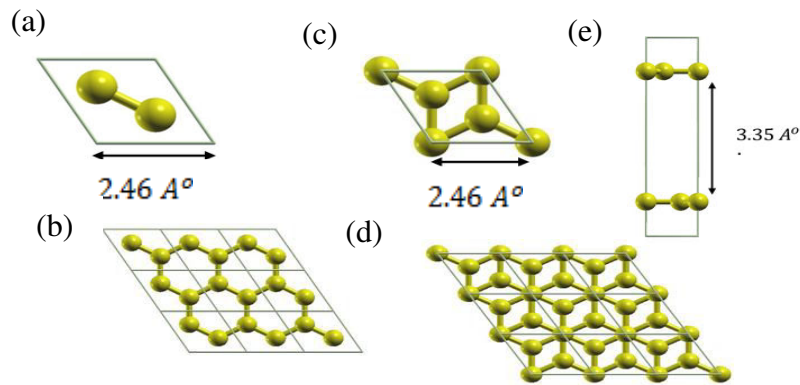

**Figure 3:** (a) Unit cell of graphene monolayer. (b) Top view of 2x2 graphene monolayer (c) Unit cell of graphene bilayer (d) Top view of 3x3 graphene bilayer (e) Side view of graphene bilayers.

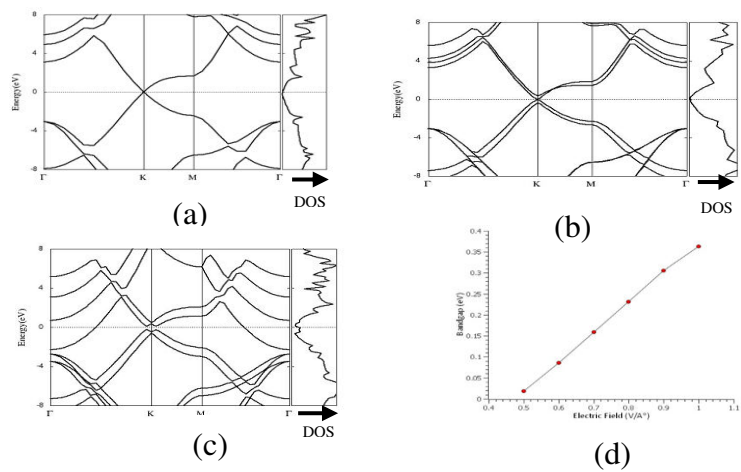

**Figure 4:** Band diagram and DOS for graphene (a) monolayer unit cell (b) bilayer unit cell (c) graphene bilayer with external EF. (d) Band gap vs external EF in bilayer

### 3.2 Silicene

In the DFT calculation used optimized parameters; lattice constants $a = b = 3.891061198\ A^o$, buckled height of Si= 0.44 A$^o$, cut off energy =30 $Ry$, and K-mesh 30 30 1. The band structure and DOS plots per unit cell are agreed with the available literature resources[15].

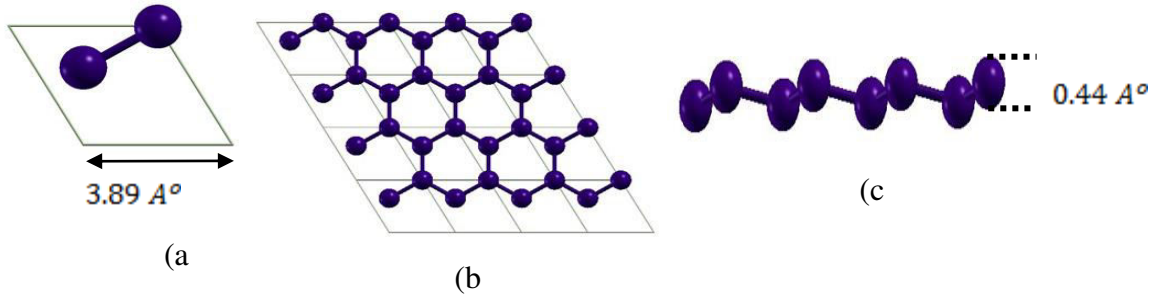

**Figure 5:** (a) Unit cell of silicene monolayer (b) 3x3 silicene monolayer (c) side view of silicene

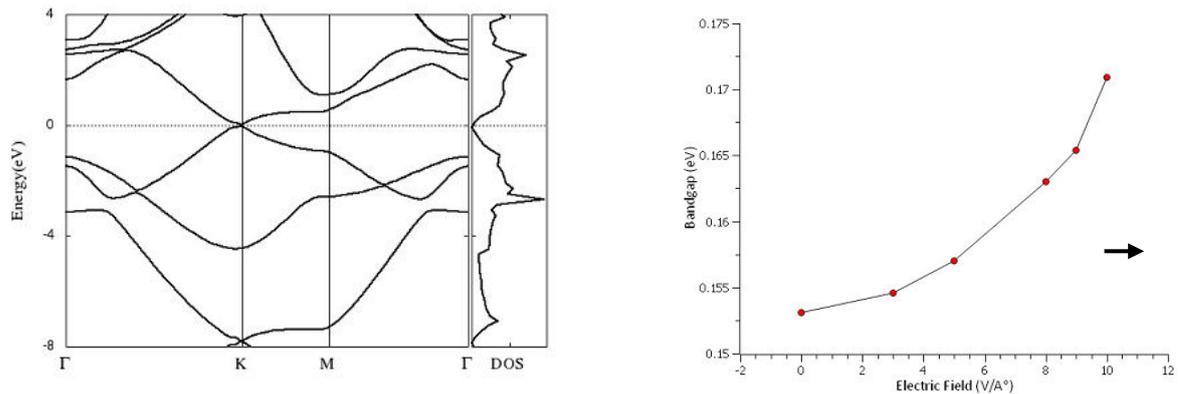

**Figure 6:** (a) Band structure and DOS (b) Bang gap vs EF for silicene unit cell.

### 1.3 Si$_2$C$_6$

In this calculation considered a unit cell consists of two silicon and six carbons atom named as Si$_2$C$_6$. (Figure 7-(a)). Lattice constants are $a = b = 4.24\ A^o$, buckle height for Si = 0.44 A$^o$, cut off energy =60 Ry, Interlayer distance = 4.4 A$^o$, k-mesh 30 30 1.
The band structure diagram for Si$_2$C$_6$ is shown in figure 8-(a). The hybrid structure mostly, favorable for forming semiconductors, but here it implies metallic properties at the Γ point, as the valence and the conduction bands overlap at this point. Additionally, it shows an n-type semiconductor gap at the K-point. The fermi level reaching to the conduction band at this point. In the structure feed in to the calculation, the graphene and silicene layers are located/oriented based on the two different symmetry axes as shown in figure7-(b) in red color dashed lines. Therefore, this may have leads to form of metallic properties at the Γ point, which is a deviation from the behavior reported on the literature for the two layers oriented in same symmetry line for other Gra/si hybrid structures[13] with respect to K-point of individual graphene and silicene. This implies creating semiconductor gaps depend on the rotation between graphene and silicene layers.

46

Proceedings of the Technical Sessions, 36 (2020)
Institute of Physics – Sri Lanka

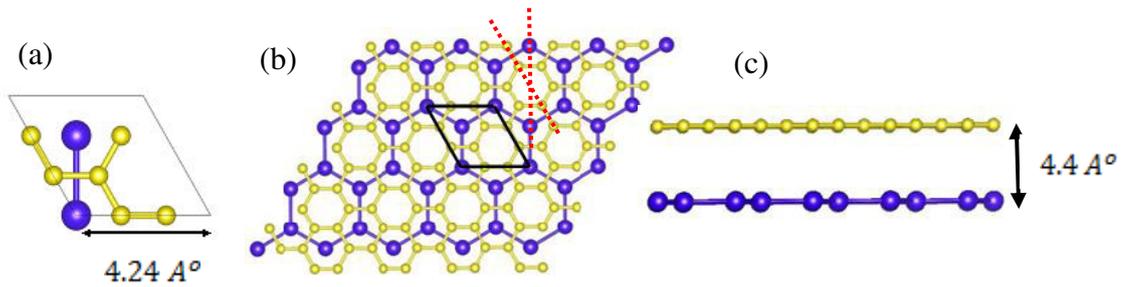

**Figure 7:** (a) Unit cell of $Si_2C_6$ (b) 4x4 $Si_2C_6$ (c) Side view of $Si_2C_6$.

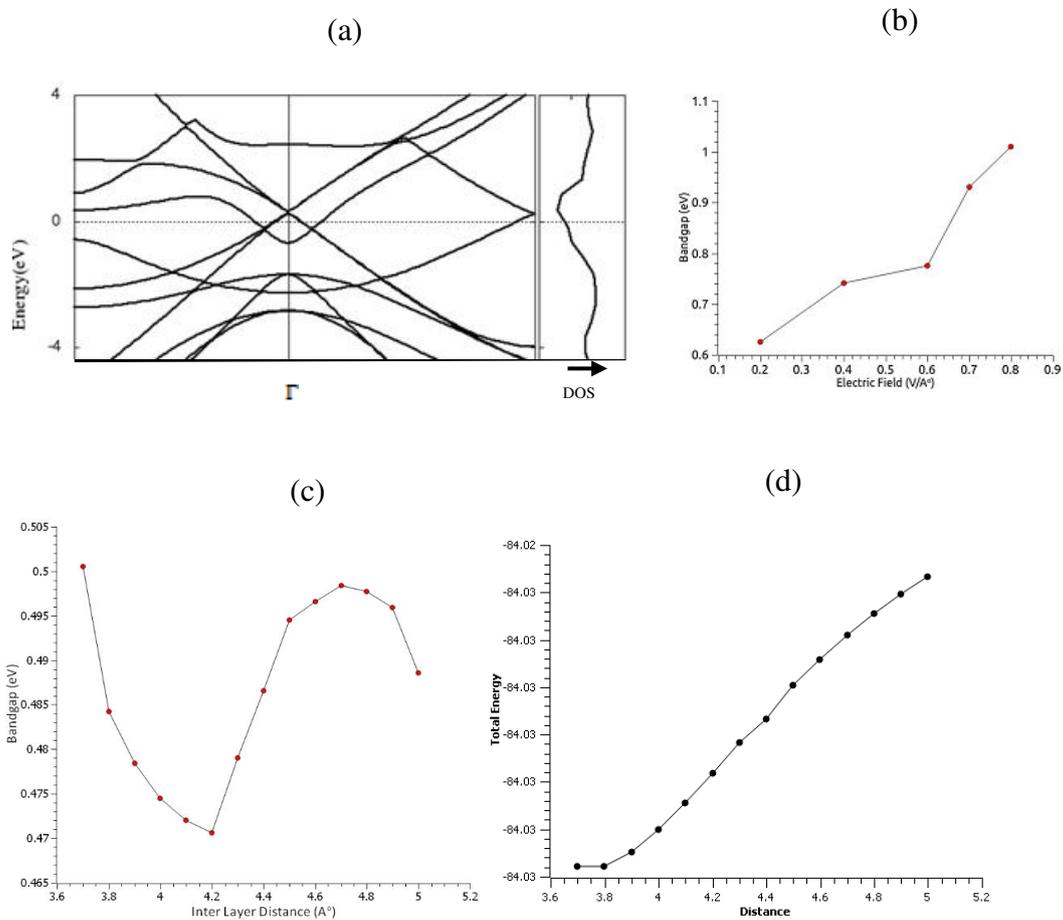

**Figure 8:** (a) Band diagram (b) band gap vs EF (c) Band gap (HOMO-LUMO difference) vs inter layer distance (d) Total energy vs interlayer distance for $Si_2C_6$

Interfacial distance is another crucial parameter for tuning the band gap and explain the electronic properties of these hybrid structures. The figure 8-(c) shows how varies the band gap (HOM LUMO difference) for different distance between (*d*) silicene and graphene layers. First, the energy gap is decreased upto 4.2 $A°$ which is implied silicene and graphene will tend to show their intrinsic properties. Further, d increases energy gap increases and decreases accordingly.The interfacial distance effects and critical point of 4.2 $A^o$ discussed in the literature for other types of Gra/si hybrids[13]. The smooth curve of total energy vs d in figure 8-(d) further ensured the band gap behavior with respect to interfacial distance. The interfacial distance leads to the form of p-n junctions in this type of hybrid structures.

As some of the future work worthy to check band structure diagrams for different interfacial distances to understand the formation of p-n junction within the hybrid structure of $Si_2C_6$.

## 4. CONCLUSION

Graphene and silicene are attractive 2D materials, which show unusual electrical and physical properties. Both materials follow hexagonal symmetry, gapless Dirac point and high mobility in charge carriers. The formation of Gra/si hybrid structures are very attractive in the formation of semiconductor junctions due to changing their interfacial distances. This semiconducting junction gap can be tuned applying transverse EF for all the configurations. Hybrid structures of $Si_nC_m$ combinations show different properties depend on their symmetry, ratio of n:m, and interfacial distances. The preliminary study is done with a $Si_2C_6$ structure, and the rotation in the symmetry axis of the silicene layer and graphene layer was introduced in to the calculation. The band diagram of this hybrid structure implies an internal Schottky (metal semiconductor) junction, with metallic properties at the $\Gamma$ −point and n-type properties at the K-point. Additionally, a band gap minima was observed at the interlayer distance of ~4.2 $A^o$. Further, studies are required to clarify the reasons of forming metallic properties at the $\Gamma$ −point, in the hybrid structure.

**ACKNOWLEDGEMENT**

I would like to take this opportunity to thank our collabrator Prof. Gayanath W. Fernando, University of Connecticut, USA for introducing our research group to Prof. Hai-Qing Lin (HQL), Director, Beijing Computational Science Research Center (CSRC). I am very much grateful to thank Prof. HQL is facilitating supercomputer access at CSRC.

**REFERENCE**

[1] Novoselov, K.S., A.K. Geim, S.V. Morozov, D. Jiang, Y. Zhang, S.V. Dubonos, I.V. Grigorieva, A.A. Firsov, *Electric Field Effect in Atomically Thin Carbon Films*, *Science,* 306 (2004) p. 666.
[2] Partoens, B., F.M. Peeters, *From graphene to graphite: Electronic structure around the $K$ point*, *Physical Review B,* 74 (2006) p. 075404.
[3] Takeda, K., K. Shiraishi, *Theoretical possibility of stage corrugation in Si and Ge analogs of graphite*, *Physical Review B,* 50 (1994) pp. 14916-14922.
[4] Cahangirov, S., M. Topsakal, E. Aktürk, H. Şahin, S. Ciraci, *Two- and One-Dimensional Honeycomb Structures of Silicon and Germanium*, *Phys. Rev. Lett.,* 102 (2009) p. 236804.
[5] Castro, E.V., K.S. Novoselov, S.V. Morozov, N.M.R. Peres, J.M.B.L. dos Santos, J. Nilsson, F. Guinea, A.K. Geim, A.H.C. Neto, *Biased Bilayer Graphene: Semiconductor with a Gap Tunable by the Electric Field Effect*, *Phys. Rev. Lett.,* 99 (2007) p. 216802.

[6] Zhou, S.Y., G.H. Gweon, A.V. Fedorov, P.N. First, W.A. de Heer, D.H. Lee, F. Guinea, A.H. Castro Neto, A. Lanzara, *Substrate-induced bandgap opening in epitaxial graphene*, Nat. Mater., 6 (2007) pp. 770-775.

[7] Elias, D.C., R.R. Nair, T.M.G. Mohiuddin, S.V. Morozov, P. Blake, M.P. Halsall, A.C. Ferrari, D.W. Boukhvalov, M.I. Katsnelson, A.K. Geim, K.S. Novoselov, *Control of Graphene's Properties by Reversible Hydrogenation: Evidence for Graphane*, Science, 323 (2009) p. 610.

[8] Berger, C., Z. Song, T. Li, X. Li, A.Y. Ogbazghi, R. Feng, Z. Dai, A.N. Marchenkov, E.H. Conrad, P.N. First, W.A. de Heer, *Ultrathin Epitaxial Graphite: 2D Electron Gas Properties and a Route toward Graphene-based Nanoelectronics*, The Journal of Physical Chemistry B, 108 (2004) pp. 19912-19916.

[9] Ball, P., *Graphene finds its place*, Nat. Mater., 13 (2014) pp. 226-226.

[10] Hill, E.W., A. Vijayaragahvan, K. Novoselov, *Graphene Sensors*, IEEE Sens. J., 11 (2011) pp. 3161-3170.

[11] Kim, K.S., Y. Zhao, H. Jang, S.Y. Lee, J.M. Kim, K.S. Kim, J.-H. Ahn, P. Kim, J.-Y. Choi, B.H. Hong, *Large-scale pattern growth of graphene films for stretchable transparent electrodes*, Nature, 457 (2009) pp. 706-710.

[12] McCann, E., V.I. Fal'ko, *Landau-Level Degeneracy and Quantum Hall Effect in a Graphite Bilayer*, Phys. Rev. Lett., 96 (2006) p. 086805.

[13] Hu, W., Z. Li, J. Yang, *Structural, electronic, and optical properties of hybrid silicene and graphene nanocomposite*, The Journal of Chemical Physics, 139 (2013) p. 154704.

[14] Lee, K.W., C.E. Lee, *Extreme sensitivity of the electric-field-induced band gap to the electronic topological transition in sliding bilayer graphene*, Sci. Rep., 5 (2015) p. 17490.

[15] Miró, P., M. Audiffred, T. Heine, *An atlas of two-dimensional materials*, Chem. Soc. Rev., 43 (2014) pp. 6537-6554.